\documentclass[twocolumn,showpacs,preprintnumbers,amsmath,amssymb]{revtex4}

\usepackage{graphicx}
\usepackage{dcolumn}
\usepackage{bm}

\begin{document}

\preprint{}

\title{Applicability of Rydberg atoms to quantum computers}

\author{Igor~I.~Ryabtsev}
  \email{ryabtsev@isp.nsc.ru}
\author{Denis~B.~Tretyakov}
\author{Ilya~I.~Beterov}
\affiliation{Institute of Semiconductor Physics\\ Prospekt
Lavrentyeva 13, 630090 Novosibirsk, Russia }

\date{January 30, 2004; revised May 14, 2004}

\begin{abstract}
Applicability of Rydberg atoms to quantum computers is examined
from experimental point of view. In many theoretical proposals
appeared recently, excitation of atoms into highly excited Rydberg
states was considered as a way to achieve quantum entanglement in
cold atomic ensembles via dipole-dipole interaction that could be
strong for Rydberg atoms. Appropriate conditions to realize a
conditional quantum phase gate have been analyzed. We also present
the results of modeling experiments on microwave spectroscopy of
single- and multi-atom excitations at the one-photon 37S$_{1/2}$$
\to $37P$_{1/2}$ and two-photon 37S$_{1/2}$$ \to$38S$_{1/2}$
transitions in an ensemble of a few sodium Rydberg atoms. The
microwave spectra were investigated for various final states of
the ensemble initially prepared in its ground state. The quantum
NOT operation with single atoms was found to be affected by the
Doppler effect and fluctuations of the microwave field. The
spectrum of full excitation of several Rydberg atoms was much
narrower than that of a single atom. This effect might be useful
for the high-resolution spectroscopy. The results may be also
applied to the studies on collective laser excitation of
ground-state atoms aiming to realize quantum gates.
\end{abstract}

 \maketitle

\section{INTRODUCTION}

Quantum Computers (QC) are of great interest last years \cite{1}.
Numerous proposals have been made on the practical realization of
QC based on the NMR in complex molecules, cold ions in
electrostatic traps, QED systems with high-Q cavities, and cold
atoms trapped in optical lattices. The molecules, ions, and QED
systems are well suited for modeling experiments demonstrating
simple quantum operations on several qubits, but are difficult to
scale to larger number of qubits required for a realistic QC
\cite{2}. More promising experiments could be made with the cold
trapped neutral atoms having advantages of low decoherence rate,
scalability to large number of qubits, and ability to control
their states by individual addressing of resonant optical pulses.

Atoms in optical lattices are usually trapped in two or three
dimensions at the antinodes of the standing light waves  \cite{3}.
In other schemes implemented recently, atoms were trapped in the
focuses of a two-dimensional array of microlenses spaced by 125
$\mu $m \cite{4}, or in the focuses of two laser beams spaced by a
few micrometers \cite{5}. Ideally, each antinode should contain a
single trapped atom representing a single qubit. Deterministic
loading of single atoms into the focus is a delicate problem that
was studied in Ref. \cite{6}.

The next stage is the operation on single or two qubits. Quantum
computation can be performed using a sequence of single-qubit
(NOT) and two-qubit (Controlled-NOT, or CNOT) operations in an
ensemble of qubits \cite{7}. In the case of alkali atoms, a
two-level qubit is represented by two hyperfine sublevels of a
ground S state. The NOT operator is implemented by Raman
transitions between the sublevels, induced by a two-frequency
laser pulse that inverts quantum state of qubit. The CNOT operator
is much more difficult to realize, since it corresponds to a
quantum nondemolition measurement of one ("control") qubit
interacting with another ("target") qubit, and after its action
the two qubits should be in an entangled state. The main
experimental problem is thus to make entangled any of two atoms in
the lattice, or at least any of two neighboring atoms.

Recently, the dipole-dipole interaction was proposed for use to
entangle neutral alkali atoms in lattices or in cold atomic
ensembles \cite{8}. This interaction, however, is weak for the
ground-state or low-excited atoms. In order to make the
interaction stronger, it was proposed in Refs. \cite{9,10} to
excite atoms for a while into the \textit{nL} Rydberg states with
\textit{n$\gg$}1, and after some manipulations bring them back to
ground state. Large atomic sizes (growing as \textit{n}$^{2}$) and
correspondingly large dipole moments of Rydberg atoms \cite{11}
strongly enhance the dipole-dipole interaction. This interaction
changes the energies of Rydberg states, and may be viewed as a
Stark shift in an atom placed in the dipole electric fields of
other atoms. The energy of this atom thus depends on the states of
neighboring atoms (whether they are in a Rydberg or in a ground
state). Entangled states may be generated using a sequence of
optical pulses applied to chosen qubits \cite{9,12,13} or using
dipole-induced changes in the spectra of collective excitations in
the ensemble of Rydberg atoms (``dipole blockade'') \cite{10,14}.

Similar considerations were also made for the van der Waals (vdW)
interaction between Rydberg atoms \cite{15}. However, this
interaction could be strong only for high Rydberg states and short
interatomic distances comparable with the size of Rydberg atoms.
In comparison with dipole-dipole interaction, it is advantageous
due to independence of the orientation of dipoles, but seems to be
impractical from its weakness. Nevertheless, experimental studies
of the vdW interaction are of interest since it may affect the
laser excitation of Rydberg states. The first experimental
observation of vdW for Rydberg atoms in a dense atomic beam was
reported in \cite{16}, and some experimental indications of the
local vdW blockade in the ensemble of cold Rydberg atoms were
reported recently \cite{17}.

Shifts and broadenings due to dipole-dipole interactions in large
ensembles of cold Rydberg atoms in magneto-optical traps were
observed for the Stark resonances in resonant collisions
\cite{18}, and recently for the microwave transitions between
Rydberg states \cite{19}. However, experimental observation of
dipole-induced frequency shifts of transitions for two Rydberg
atoms or dipole blockade for a small ensemble have not been
reported yet. Such experiments meet many technical difficulties,
require a dense optical lattice with a detection system for
Rydberg atoms and powerful narrow-band lasers for fast excitation
of Rydberg states from the ground state.

On the other hand, some modeling experiments with low number of
atoms may be performed in an atomic beam with microwave
transitions between neighboring Rydberg states. Such experiments
can reveal important features of the multi-particle excitation
spectra of Rydberg atoms.

In this paper we discuss the applicability of Rydberg atoms to QC
from a practical point of view, and analyze conditions required to
implement proposed schemes in practice. We also present the
results of modeling experiments on the microwave spectroscopy of
multi-atom excitations of sodium Rydberg atoms.

\section{EXPERIMENTAL LIMITATIONS ON QC WITH RYDBERG ATOMS}

\subsection{Principal quantum numbers: limitations by collisions}

The dipole moment of an atom is proportional to the average radius
\textit{r} of the electron orbit. We have to choose \textit{r} as
large as possible in order to use advantages of Rydberg atoms. In
a hydrogen atom, \textit{r} is given by the formula \cite{20}

\begin{equation}
\label{eq1} r = \frac{{1}}{{2}}a_{0} \left\{ {\;3n^{2} - L\left(
{L + 1} \right)} \right\},
\end{equation}

\noindent where \textit{a}$_{0}$ is the Bohr radius. \textit{r}
has close value also for the Rydberg states of hydrogen-like
alkali atoms. A short distance \textit{R} between neighboring
atoms in the lattice imposes obvious limitation on the maximal
value of \textit{r}: \textit{R} must exceed \textit{r} by an order
of magnitude to avoid collisions between electrons of close
Rydberg atoms. Collisions have large cross-sections (comparable
with atomic size), and may cause fast decoherence of Rydberg
states (e.g., mixing of degenerate magnetic sublevels or
stochastic jumps in the phases of wave-functions). Assuming
\textit{r}$<$\textit{R}/10, the maximum principal quantum number
\textit{n}$_{max}$ of Rydberg atoms in the lattice is thus limited
by

\begin{equation}
\label{eq2}
n_{max} \sim \sqrt {\frac{{R}}{{15a_{0}} } +
\frac{{L\left( {L + 1} \right)}}{{3}}} .
\end{equation}

In the optical lattices of alkali atoms, \textit{R} can vary from
$\sim $0.5 $\mu $m (in red-detuned dipole traps) to $\sim $5 $\mu
$m (in quasi-electrostatic traps with a CO$_{2}$-laser) \cite{3},
or to 125 $\mu $m for microlenses array \cite{4}. The spacing 0.5
$\mu $m is too small for reliable individual optical access to
each qubit without disturbing the neighboring qubits, while 125
$\mu $m is too long for effective dipole-dipole interaction (see
the next subsection). The 5 $\mu$m value seems to be optimal for
both of optical access and compatibility with CO$_{2}$-laser traps
having an advantage of extremely low scattering rate of photons.

From Eq.(\ref{eq2}) for S-states (\textit{L}=0) one obtains
\textit{n}$_{max}\sim $70 at \textit{R}$\sim $5 $\mu $m. For
circular Rydberg states \cite{11} with maximum possible
\textit{L}=(\textit{n}-1), the estimate is \textit{n}$_{max}\sim
$90 at \textit{R}$\sim $5 $\mu $m. Excitation of circular states,
however, requires much time and sophisticated combination of the
optical and microwave pulses in the magnetic and electric fields.
One can conclude that in practice low-\textit{L} states with
$n<$70 should be chosen, although higher \textit{n} would result
in a stronger dipole interaction, and experiments with
\textit{n}$\sim $500 have been already reported \cite{21}.

For further considerations we take \textit{n$<$}70 and \textit{R$
\simeq $}5 $\mu $m as preliminary parameters satisfying the
following experimental conditions: (i) \textit{R}$ \simeq $5 $\mu
$m is a spacing that provides free optical access to each qubit
and compatibility with CO$_{2}$-laser lattices; (ii) \textit{n$
\simeq $}70 is a maximal principal quantum number at given
\textit{R} that ensures absence of collisions between neighboring
Rydberg atoms.

\subsection{Dipole-dipole interaction and time of quantum operation}

Now we have to analyze the dipole-dipole interaction at the
\textit{n} and \textit{R} values specified above. Consider the
interaction between two identical Rydberg atoms \textit{a} and
\textit{b} spaced by a distance \textit{R} (Fig.1). These two
atoms may be viewed as a quasi-molecule whose energy eigenstates
should be determined in a molecular wavefunctions basis $\left|
{\Psi _{a} \,\Psi _{b}} \right\rangle $, as it was made in Refs.
\cite{8,9,10}. For example, if we choose for the consideration two
close nondegenerate Rydberg states of opposite parity, $\left| 1
\right\rangle $ and $\left| 2 \right\rangle $, molecular states
$\left| 12 \right\rangle $ and $\left| 21 \right\rangle $ are
degenerate in energy and experience strong dipole-dipole
interaction. The eigenstates of the quasi-molecule are readily
found by solving the secular equation with the interaction
Hamiltonian

\begin{equation}
\label{eq3} \hat {H}_{dd} = \frac{{1}}{{4\pi \varepsilon _{0}}
}\left[ {\frac{{\hat {\mathbf{d}}_{a} \hat {\mathbf{d}}_{b}}
}{{R^{3}}} - \frac{{3\,\left( {\hat {\mathbf{d}}_{a} \mathbf{R}}
\right)\left( {\hat {\mathbf{d}}_{b} \mathbf{R}}
\right)}}{{R^{5}}}} \right],
\end{equation}

\noindent where $\hat{\mathbf{d}}_{a}$ and $\hat{\mathbf{d}}_{b}$
are dipole moment operators of atoms \textit{a} and \textit{b},
$\mathbf{R}$ is a vector connecting two atoms, and $\varepsilon
_{0}$ is the dielectric constant. The solution shows that
degenerate states of the quasi-molecule are split by the
dipole-dipole interaction to two sublevels with symmetric
(superradiant) and antisymmetric (subradiant) wave functions

\begin{equation}
\label{eq4}
{\begin{array}{*{20}c}
 {\left| { +}  \right\rangle = \frac{{1}}{{\sqrt {2}} }\left( {\left|
 12 \right\rangle + \left|  21 \right\rangle}  \right)}, \hfill \\
 {\left| { -}  \right\rangle = \frac{{1}}{{\sqrt {2}} }\left( {\left|
 12 \right\rangle - \left|  21 \right\rangle}  \right)}.
\hfill \\
\end{array}}
\end{equation}

\begin{figure}
\includegraphics[scale=1]{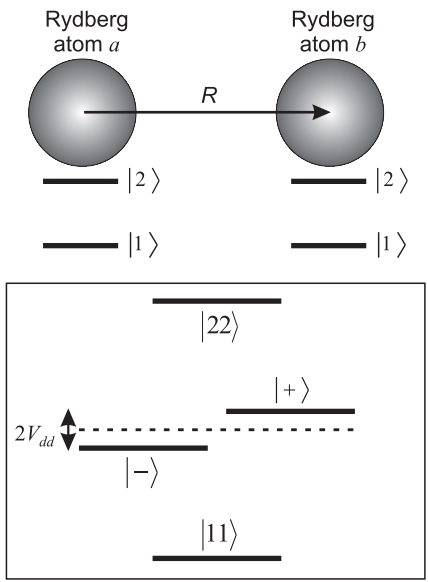}
\caption{\label{Fig1}A model for the dipole-dipole interaction
between two Rydberg atoms. States $\left|  {1} \right\rangle $ and
$\left|  {2} \right\rangle $ are two neighboring Rydberg levels.
States $\left|  11 \right\rangle ,\,\left| { - } \right\rangle
,\;\left| { +} \right\rangle ,\left| 22 \right\rangle $ are
eigenstates of a composite system of two interacting atoms.}
\end{figure}

If the quantization axis \textit{z} is chosen along $\mathbf{R}$,
and states $\left| 1 \right\rangle $ and $\left| 2 \right\rangle $
have identical momentum projections that are bound only by the
$\hat{d}_{z}$ components of dipole operators (e.g., particular
Zeeman sublevels of states $\left| 1 \right\rangle $ and $\left| 2
\right\rangle $ whose degeneracy is lifted by a magnetic field),
the energy shifts of these states are

\begin{equation}
\label{eq5}
V_{dd} = \pm 2d_{z}^{2} /\left( {4\pi \varepsilon _{0}
R^{3}} \right),
\end{equation}

\noindent where \textit{d}$_{z}$ is \textit{z} component of the
dipole matrix element of the 1$ \to $2 transition.

Addressing short individual laser pulses to the atoms, we can
initially excite atom \textit{a} into state $\left| 1
\right\rangle $ and atom \textit{b} into state $\left| 2
\right\rangle $. Then atoms start to interact, and the time
evolution of the wavefunction of the quasimolecule is following:

\begin{equation}
\label{eq6} \Psi \left( {t} \right) = \left| 12 \right\rangle
\cos\left( {V_{dd} t/\hbar}  \right) - i\left| 21 \right\rangle
\sin\left( {V_{dd} t/\hbar} \right),
\end{equation}

\noindent i.e., population oscillates between states $\left| 12
\right\rangle $ and $\left| 21 \right\rangle $ with a frequency
$\left( {2\,V_{dd} /\hbar} \right)$.

This process may be considered as an exchange of virtual photons
between two atoms. It was already observed experimentally when two
Rb atoms in circular Rydberg states with \textit{n}=50 and 51 were
passed through a high-Q superconducting microwave cavity with
variable detuning from resonance transition between these states
\cite{22}. The cavity strongly enhanced the interaction so that
photon exchange occurred at \textit{R}$\sim $1 mm on a time scale
of $\sim $100 $\mu $s. Huge gain of the dipole-dipole interaction
makes high-Q cavities attractive for modeling experiments on QC
employing atoms passed through the cavity. However, in more
realistic QC with cold trapped atoms it would complicate the
loading of the lattice and optical access to atoms, and require
special arrangement for cavity cooling to the temperatures $\sim
$1 K. Cooling is needed to achieve the superconductivity and, what
is more important, to suppress the photons of blackbody radiation
(BBR) that induce uncontrolled transitions between Rydberg states
\cite{11}. The cavity significantly increases rates of
BBR-transitions, as we have demonstrated in \cite{23}, although
usage of detuned cavity in \cite{22} made atoms less sensitive to
BBR. Further we will analyze the interaction in a free space,
since practical application of a cavity for enhancement of
dipole-dipole interaction is ambiguous.

From Eq.(\ref{eq6}) one finds that in a free space the population
returns back to state $\left| 12 \right\rangle $ after the time

\begin{equation}
\label{eq7}
T = \pi \,\hbar /V_{dd} ,
\end{equation}

\noindent while the wavefunction changes its sign, or obtains a
phase shift $\pi $. Such evolution is well suited for realization
of a conditional Quantum Phase Gate (QPG) proposed in \cite{24}
and realized in a microwave cavity experiment with Rb Rydberg
atoms \cite{25}. QPG is of great importance since it may be
consequently converted to CNOT gate with additional rotation of
one of qubits. QPG was also analyzed theoretically in \cite{12}
for two Rb Rydberg atoms trapped in close laser focuses, as in
experiment \cite{5}. A scheme was considered in which two atoms
were initially excited to the same 42S Rydberg state, lying almost
in the middle between two neighboring states 41P and 42P. Similar
situation was also employed in proposals on the dipole blockade at
multi-atom excitations in the atomic ensemble, but in this case an
electric field was required to tune transitions to exact resonance
\cite{10,14}.

We propose another scheme that uses only two \textit{n}S and
\textit{n}P Rydberg states in two alkali atoms \textit{a} and
\textit{b} (Fig.2), and no Stark tuning is needed. The ground
S-state has two hyperfine sublevels F=1 and F=2 (for example, in
$^{23}$Na and $^{87}$Rb atoms) which represent two qubit states
$\left| {0} \right\rangle $ and $\left| {1} \right\rangle $. Atom
\textit{a} is excited from state $\left| {1} \right\rangle $ to
\textit{n}S Rydberg state by a $\pi $/2 laser pulse via absorption
of two photons, while atom \textit{b} is excited from state
$\left| {1} \right\rangle $ to \textit{n}P Rydberg state by a $\pi
$/2 laser pulse via one-photon absorption. [Expressions describing
$\pi $/2 pulses for the one- and two-photon transitions will be
given below in Eqs.(\ref{eq11}), (\ref{eq12}), and (\ref{eq13})].
Atoms are returned back to the ground state by the subsequent
3$\pi $/2 laser pulses. The final phase of the composite
wavefunction changes by $\pi $ if the dipole-dipole interaction
occurs between exciting and de-exciting laser pulses during time
\textit{T} of Eq.(\ref{eq7}). In any other cases of initial states
of qubits there will be no phase shifts, as is required for QPG.
The laser pulses should be strong and short enough to make
excitation and de-excitation insensitive to dipole-induced energy
shifts. Eq.(\ref{eq7}) determines minimum time of QPG.

\begin{figure}
\includegraphics[scale=1]{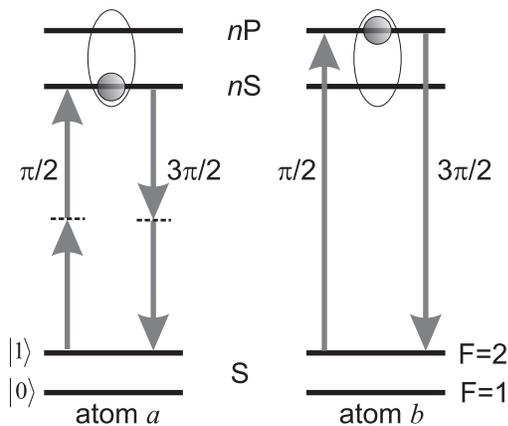}
\caption{\label{Fig2} Scheme of QPG operation with two Rydberg
atoms. The qubit states $\left| {0} \right\rangle $ and $\left|
{1} \right\rangle $ are hyperfine sublevels of the ground state.
Atom \textit{a} is excited to the \textit{n}S state via two-photon
absorption, and atom \textit{b} is excited to the \textit{n}P
state via one-photon absorption of $\pi $/2 laser pulses. After
dipole-dipole interaction they are returned back to the ground
state by 3$\pi $/2 laser pulses.}
\end{figure}

We can also analyze an opportunity to observe the frequency shifts
of microwave transitions between Rydberg states for modeling the
dipole blockade effect \cite{10}. For instance, we may apply a
microwave radiation to both atoms initially excited to \textit{n}S
state. Recording the absorption spectrum, we should see a shift in
the frequency of single-atom excitation of two atoms at the
transition $\left| \mathrm{SS} \right\rangle \to \left| { +}
\right\rangle $, in comparison with unperturbed frequency of the
\textit{n}S$ \to $\textit{n}P transition in a single Rydberg atom.
The frequency of the transition $\left| \mathrm{SS} \right\rangle
\to \left| { -}  \right\rangle $ is also shifted, but probability
of this transition is zero due to antisymmetric wave function of
$\left| { -}  \right\rangle $ state.

Dipole-dipole interaction thus affects the frequencies of
transitions in an atomic ensemble. If we apply a pulse of
microwave radiation at the unperturbed frequency of the
\textit{n}S$ \to $\textit{n}P transition, and the shift is large
enough, the final state of Rydberg atom will depend on whether the
neighboring atoms are in Rydberg or in ground states. A necessary
condition is that the frequency shift $(V_{dd} /h)$ must exceed
spectral width of the pulsed radiation that is ultimately limited
by its duration $\tau $:

\begin{equation}
\label{eq8}
V_{dd} /h > > 1/\tau .
\end{equation}

Fast QC requires \textit{T} from Eq.(\ref{eq7}) or $\tau $ from
Eq.(\ref{eq8}) to be as short as possible. Eq.(\ref{eq7}) is more
appropriate to satisfy this requirement, and we see that the QPG
gate needs much smaller \textit{V}$_{dd}$. In order to find
\textit{T} from Eq.(\ref{eq7}), we estimate achievable values of
\textit{V}$_{dd}$ for \textit{R}$\sim $5 $\mu $m and
\textit{n}$\sim $50. The radial part \textit{d} of dipole moments
of transitions (\textit{nL$ \to $nL$ \pm $}1) between close
Rydberg states is roughly estimated as

\begin{equation}
\label{eq9}
d\sim e\;a_{0} \;n^{2},
\end{equation}

\noindent where \textit{e} is the electron charge. Precise
calculations can be made using numerical integration of the
Schr$\ddot{\mathrm{o}}$dinger's equation described in \cite{26},
or with analytical expression and tables derived in \cite{27} for
arbitrary effective quantum numbers
\textit{n}$_{eff}$=\textit{n}-$\delta _{L} $ ($\delta _{L} $ is a
quantum defect). As an example, we take the 50S$_{1/2} \to
$50P$_{1/2}$ microwave transition in Na at 27.7 GHz and in Rb at
30.0 GHz. Numerically calculated radial part of the dipole moment
is 2690 a.u. in Na and 2550 a.u. in Rb, in good agreement with
Eq.(\ref{eq9}). With angular part being 1/3, one obtains $(V_{dd}
/h)\sim $10 MHz and \textit{T}$\sim $50 ns. We see that in
principle the QPG operation based on dipole-dipole interaction of
Rydberg atoms can be rather fast. However, as we will see below,
achievable speed of laser excitation of Rydberg states imposes
serious practical limitations.

\subsection{Laser excitation}

Laser pulses exciting and de-exciting Rydberg states must be much
shorter than \textit{T} to provide sharp switching and ensure the
validity of Eq.(\ref{eq6}). At \textit{T}$\sim $50 ns the pulses
should be of $\tau \sim $1-5 ns duration. The use of short pulses
encounters two serious obstacles.

First, the probabilities of single-photon transitions from
low-excited to Rydberg states fall as \textit{n}$^{-3}$ \cite{11}.
This is a forcible argument not to use very high Rydberg states:
short laser pulses would require too high laser power to generate,
e.g., 3$\pi $/2 pulses for the de-excitation of Rydberg states to
ground states. For example, the calculated laser intensity
required for a 3$\pi $/2 pulse of 5 ns duration is 120 MW/cm$^{2}$
for the (3S$_{1/2}$ F=2 M$_F$=2)$\to$(50P$_{3/2}$ F=3 M$_F$=3)
transition at 240 nm in Na, and 30 MW/cm$^{2}$ for the (5S$_{1/2}$
F=2 M$_F$=2)$\to$(50P$_{3/2}$ F=3 M$_F$=3) transition at 300 nm in
Rb.  Even if laser radiation is focused to a spot of $\sim $1 $\mu
$m size, the laser powers should be $\sim $1 W for Na and $\sim
$0.25 W for Rb. Also important is that each atom in the lattice
should be controlled by its own laser beam, so the total laser
power required for the lattice, say, of 100 atoms would be
extremely large. For UV radiation at 250-300 nm such powers are
difficult to obtain from frequency doubled or tripled cw lasers
with external modulators. At the same time, powerful pulsed lasers
can not provide the frequency and amplitude stability required for
reliable and reproducible quantum operations.

Second, 5 ns pulses have $\sim $200 MHz spectral width that is not
enough for spectral resolution of transitions between particular
sublevels of the ground and Rydberg states. For example, the
calculated fine structure intervals of the 50P state are 45 MHz in
Na and 819 MHz in Rb. The resolution of $\sim $20 MHz is thus
required for excitation of particular sublevels of Rydberg states.

Both of obstacles specified above may be overcome simply by
increasing the duration of laser pulses to $\sim $50 ns. However,
since time \textit{T} of dipole-dipole interaction defined by
Eq.(\ref{eq7}) should be an order of magnitude longer than laser
pulses, \textit{T} has to be increased to $\sim $500 ns by
lowering $\left( {V_{dd} /h} \right)$ to $\sim $1 MHz. The
decrease may be achieved by lowering $n$ value to $\approx$30.
Laser power per atom decreases to reasonable values of $\sim $2 mW
for Na and $\sim $0.5 mW for Rb, and spectral resolution will be
improved to $\sim $20 MHz.

In general, Rydberg states with \textit{n}=25-50 in Na and Rb
atoms may be excited from the ground states by single- or
multi-photon absorption. A single-photon process requires UV laser
radiation and allows the excitation of \textit{n}P-states only (in
the presence of electric field, however, other states may be
excited due to the mixing of the zero-field wavefunctions). For Rb
atoms, UV radiation at 300 nm is required, that may be obtained by
doubling the frequency of a Rhodamine 6G dye-laser at 600 nm. On
the other hand, intense 600 nm radiation itself can excite the
\textit{n}S and \textit{n}D Rydberg states of Rb via two-photon
absorption. The scheme of Fig.2 thus may be realized with two
Rhodamine 6G dye-lasers. These lasers, however, need powerful
pumping lasers (Ar or second harmonic of Nd:YAG). For compact
inexpensive Rb experimental setups, it would be better to use
multi-step excitation schemes based on frequency stabilized
semiconductor lasers, as in \cite{22,25,28}.

For Na atoms, a two-step scheme 3S$ \to $3P$ \to
$(\textit{n}S\textit{, n}D) was often used. The first step is
excited by the Rhodamine 6G dye-laser at 589 nm. The second step
was usually excited by dye-lasers near 410 nm pumped by UV
nitrogen, eximer, or third harmonic of a Nd:YAG laser. These
pumping lasers operate in a pulsed regime at a low repetition rate
(1-100 Hz) that is too slow for QC. Now the 410 nm dye-laser can
be replaced with a second harmonic of cw Ti:sapphire laser or blue
semiconductor diode laser.

Finally, taking into account limited speed of laser excitation, we
have to specify new optimal parameters for quantum operations: the
laser pulses of $\sim $50 ns duration should be used to excite and
de-excite Rydberg atoms at reasonable laser powers; the time of
QPG operation should be increased to $\sim $500 ns by decreasing
the principal quantum number to \textit{n$ \approx $}30.

\subsection{Degeneracy of atomic states}

We note that model of Fig.2 does not take into account the
degeneracy of atomic states. It is important both for the laser
excitation and dipole-dipole interaction, since dipole moments of
transitions between various magnetic sublevels of degenerate
states differ by magnitudes and signs.

Obviously, the ground-state atoms in the optical lattice must be
optically pumped to a single magnetic sublevel prior quantum
operations. Otherwise, if the ground states are arbitrary mixtures
of magnetic sublevels, the speed of laser excitation to Rydberg
states will be different for various particular transitions, and
$\pi/2$ laser pulses can not be obtained to transfer all
populations to Rydberg states.

Accounting for the magnetic sublevels at the dipole-dipole
interaction results in several different energy levels of the
quasi-molecule of two Rydberg atoms. For instance, there are three
energy levels in the case of the simplest microwave transition
\textit{n}S$_{1/2} \to $\textit{n}P$_{1/2}$ when each state has
two magnetic sublevels $M_{J}=\pm 1/2$. The behaviour of the
quasi-molecule depends now on initial distribution of populations
over magnetic sublevels of Rydberg states. Calculations have
revealed that Eq.(\ref{eq6}) is no more valid, and the phase shift
$\pi$ of the wavefunction can not be achieved for degenerate
sublevels.

We conclude that degeneracy should be completely lifted  for both
ground and Rydberg states, e.g., by a homogeneous magnetic field.
The field should be strong enough to split the sublevels by a
value exceeding the expected energy of dipole-dipole interaction
$\left( {V_{dd} /h} \right)\sim $1 MHz in order to ensure validity
of Eq.(\ref{eq6}).

\subsection{External fields}

\textit{Magnetic field} splits degenerate sublevels due to the
Zeeman effect. The Zeeman effect in Rydberg atoms and quantum
interference of degenerate sublevels at microwave transitions in a
weak magnetic field were studied experimentally in our work
\cite{29}. The splittings are approximately 2.8 MHz/Gs for
$n$S$_{1/2}$ states and 0.93 MHz/Gs for $n$P$_{1/2}$ states. One
finds that $\sim$10 Gs field is required to split the $n$P$_{1/2}$
state by 10 MHz.

Applied magnetic field will also reduce the sensitivity of atoms
to stray magnetic fields. For example, the Earth's magnetic field
of the 0.5 Gs magnitude alone splits atomic states by $\sim $0.5
MHz. This results in a phase difference of the order of $\pi $
obtained by magnetic sublevels during $\sim $1 $\mu $s interaction
time, and in a mixing of populations due to the precession of
magnetic moment. The phase is important when optical or microwave
transitions are excited between coherently prepared degenerate
states that can interfere. Our previous experiment on microwave
Hanle effect at the 37P$_{3/2} \to $37S$_{1/2}$ microwave
transition in Na atoms demonstrated variation of the probability
by 50\% when magnetic field was varied within 20 mGs around zero
\cite{29}. In another experiment with this transition we observed
a mixing of populations due to the precession \cite{30}.
Application of a magnetic field of $\sim$10 Gs will define the
quantization axis and suppress the precessions from stray fields,
so that no additional magnetic screening will be needed.

\textit{Electric field} influences on Rydberg atoms especially
strongly. In Fig.3 a part of the numerically calculated Stark
diagram for the sodium Rydberg states near \textit{n}=36 is shown
(the calculations have been made using the diagonalization
procedure described in \cite{26}). The electric field of several
V/cm significantly changes energies of all Rydberg levels. The
hydrogen-like states with \textit{L}$>$1 experience linear Stark
effect and become completely mixed by the electric field, while
non-hydrogenic \textit{n}S and \textit{n}P states have large
quantum defects in sodium (1.348 and 0.855 correspondingly) and
experience quadratic Stark effect up to the field of 10 V/cm. The
mixing of high-moment states makes them inapplicable to QC since
any stray electric field would redistribute populations. We
conclude that in Na only S and P states would be suitable to QC
because they are stable against electric field fluctuations. An
exception is the D-state (0.015 quantum defect) that also has
quadratic Stark effect in the electric fields below 0.1 V/cm, and
is not mixed with other states in such fields.

\begin{figure}
\includegraphics[scale=1]{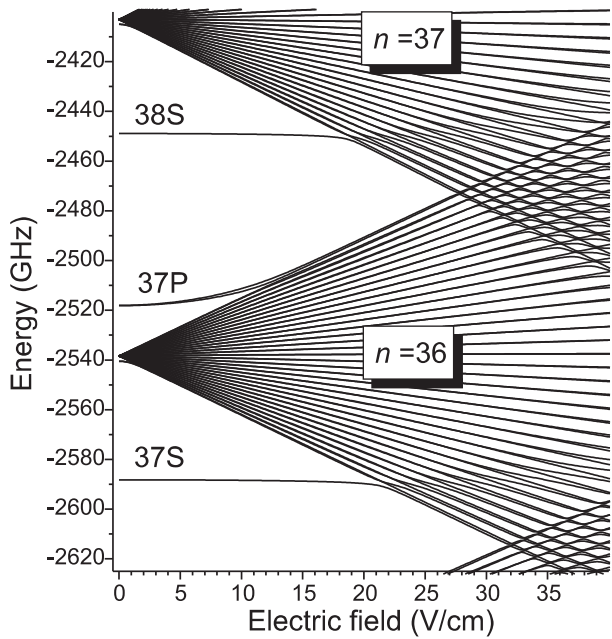}
\caption{\label{Fig3} Numerically calculated Stark diagram of the
sodium Rydberg states near \textit{n}=36.}
\end{figure}

Usage of S and P states has an advantage of ability to use
Stark-switching technique to control the interaction with cw
radiation in a coherent way, as it was demonstrated in our recent
experiment on Stark-switched Ramsey interferometry of the
37S$_{1/2} \to $37P$_{1/2}$ transition \cite{31}. The
Stark-switching can significantly simplify experiments because it
eliminates the need to use fast modulators for cw radiation, but
it would require phase corrections of quantum operations to
account for additional phase shifts from the Stark effect. Lower
Rydberg states with $n\approx$30 are less sensitive to the
electric field than states in Fig.3 since polarizabilities scale
as $n^7_{eff}$, and no electric field screening will be required
if 30S and 30P states will be used.

\textit{Blackbody radiation} (BBR) induces the microwave
transitions to nearby Rydberg states and shortens their lifetimes.
In order to decrease its influence, experiments with Rydberg atoms
are usually made with thermal shields cooled to liquid nitrogen
(77 K) or liquid helium (4 K) temperatures. This technique
requires vacuum cryostats and nontransparent copper shields that
would complicate experimental arrangements and optical access to
individual atoms in the optical lattice. We have to decide is this
cooling needed in the experiments on QC. Without the BBR, the
radiative lifetimes scale as $n^{3}_{eff}$ for the states with
\textit{L$\ll$} \textit{n} \cite{32}. In a sodium atom $\tau
$(30S)$ \approx $30 $\mu $s and $\tau $(30P)$ \approx $300 $\mu
$s. With the BBR at a room temperature 300 K, effective lifetimes
decrease to $\tau $(30S)$ \approx $20 $\mu $s and $\tau $(30P)$
\approx $30 $\mu $s. Although the decrease is significant, we see
that influence of the BBR is of small importance if the operation
time is $\sim $500 ns, and no cooled environment is needed.
However, one should keep in mind that if experiments are made with
a detection system that uses some closed metallic surrounding,
enhancement of interaction of atoms with the BBR may occur since
the surrounding works as a microwave cavity increasing spectral
density of the BBR field \cite{23}.

\subsection{Detection}

Sensitive detection of single Rydberg atoms is a crucial point of
the future experiments on QC. Only technique of the Selective
Field Ionization (SFI) \cite{11} is applicable to high Rydberg
states with \textit{n$ > $}20 since spontaneous fluorescence is
extremely low. A Rydberg atom in the \textit{nL} state can ionize
with a probability close to 1 by an electric field whose critical
strength \textit{E}$_{c}$ strongly depends on the effective
quantum number according to the formula

\begin{equation}
\label{eq10} E_{c} \approx 3.2 \cdot 10^{8}n_{eff}^{ - 4} \quad
\mathrm{V/cm.}
\end{equation}

\noindent For \textit{n}$\approx $30 one finds that
\textit{E}$_{c}\approx $400 V/cm is required.

Single electrons resulted from ionization are usually detected by
channeltron or channel-plate detectors with a probability limited
mostly by the transparency of the input electric field grid
(typically 50-80 \%). The SFI also allows to measure distribution
of populations over different Rydberg states, if a ramp of
electric field is applied.

Obvious drawback of this method is the destruction of Rydberg
atoms in the ionization process. Hence, it cannot be applied to
continuous control of Rydberg atoms in QC, and may be used only
for the preliminary tunings before quantum computation. After
loading of optical lattice with single atoms, all intermediate
operations must be performed without the SFI detection, and only
final measurements can be made with it.

Individual SFI detection of single Rydberg atoms in the optical
lattice is another technical problem. Since atoms in the lattice
will be spaced by $\sim $5 $\mu $m, position sensitive
channel-plate detectors must have proper spatial resolution to
ensure detection of each atom. Also, if individual detection of
atoms is needed, each atom should be provided with its own
electric field electrode that does not affect other atoms. Such
configuration will be difficult to realize in practice, and the
only way will be use of the common electric field plate for the
whole lattice. This means that preliminary tunings can be made
only for the whole atomic ensemble.

\subsection{Summary for experimental limitations}

$ \bullet $ Minimum distance \textit{R} between atoms in optical
lattice has to be 5 $\mu $m to provide free optical access to each
atom. This distance is compatible with conservative
quasi-electrostatic CO$_{2}$-laser traps having extremely low
scattering rate of photons.

$ \bullet $ At this $R$, principal quantum numbers \textit{n} of
Rydberg states should not exceed 70 to avoid dephasing collisions
between neighboring Rydberg atoms.

$ \bullet $ With \textit{R}$\approx$5 $\mu $m and
\textit{n}$\approx$50, maximum energy shifts \textit{V}$_{dd}$ of
Rydberg states due to the dipole-dipole interaction may reach 10
MHz, and minimum time required to perform a QPG operation is 50
ns. But in this case exciting laser pulses should be of $\sim $5
ns duration and $\sim $1 W power, even if focused to a spot of
$\sim$1 $\mu $m size.

$ \bullet $ In order to use reasonable laser powers of $\sim $1 mW
per atom and improve spectral resolution to $\sim$20 MHz at
excitation and de-excitation of Rydberg states, duration of the
laser pulses should be increased to 50 ns. The time of
dipole-dipole interaction should be also increased to $\sim $500
ns by lowering $n$ to 30.

$ \bullet $ Magnetic field of $\sim $10 Gs has to be applied to
lift the degeneracy of ground and Rydberg states, to provide
selective excitation of particular magnetic sublevels and to
suppress the precession of magnetic moment in stray magnetic
fields. Ground-state atoms should be optically pumped to a certain
magnetic sublevel prior quantum computations.

$ \bullet $ Stray electric fields should be suppressed to avoid
mixing of Rydberg states, or low-moment states having quadratic
Stark effect should be chosen, for example, \textit{n}S and
\textit{n}P states.

$ \bullet $ Suppression of the blackbody radiation at 300 K is not
required if the time of the dipole-dipole interaction is $\sim
$500 ns.

$ \bullet $ The SFI detection of Rydberg atoms may be applied only
before quantum computations for preliminary tunings  and
afterwards for final measurements.

\section{MICROWAVE SPECTROSCOPY OF MULTI-ATOM EXCITATIONS}

QC based on cold atoms in optical lattices may be realized, in
principle, in the nearest future. The above considerations impose
the limitations that require additional experimental tests. Some
experiments can be made with atoms in a thermal atomic beam. For
example, important for the dipole blockade \cite{10} spectra of
multi-atom excitations of Rydberg atoms have not been reported
yet. Even without dipole-dipole interaction, these spectra are
expected to be different from that of single-atom excitations,
since various final states of the ensemble of Rydberg atoms may be
selectively detected with the SFI technique. We studied these
spectra at microwave transitions in Na Rydberg atoms in an
effusive atomic beam.

\subsection{Experimental setup}

The sodium atoms were excited to an initial 37S$_{1/2}$\textit{}
Rydberg state by a two-step scheme 3S$_{1/2} \to $3P$_{3/2} \to
$37S$_{1/2}$ with two pulsed tunable lasers (Rhodamine 6G and the
second harmonic of Ti:sapphire) at a 5-kHz repetition rate
(Fig.4). Laser pulses were synchronized and acted simultaneously,
exciting several Rydberg atoms per pulse on average.

\begin{figure}
\includegraphics[scale=0.67]{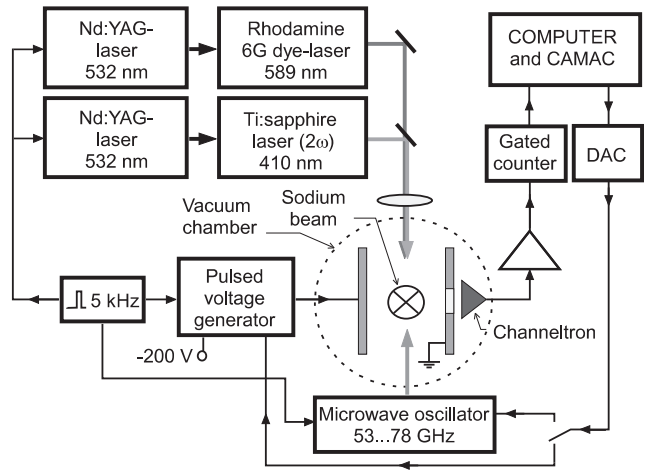}
\caption{\label{Fig4} Experimental setup. }
\end{figure}

Experiments were carried out in a vacuum chamber with a sodium
effusive atomic beam at 450-K temperature and 10$^{8}$ cm$^{-3}$
atom density in a 5$ \cdot $10$^{-5}$ cm$^{3}$ excitation volume.
No collisions and dipole-dipole interaction were expected in such
conditions since the average distance between ground-state atoms
was $\sim $10 $\mu $m, and the average distance between Rydberg
atoms was $\sim $100 $\mu $m. The laser radiation was focused
perpendicularly to the beam, so that the point of laser excitation
was localized within 0.2 mm along the beam. The Rydberg atoms were
passed between the two copper plates separated by 8.15 mm. A
pulsed voltage was applied to one of them, to form a homogeneous
electric field for the detection of atomic populations by the SFI.
The other plate had a 9-mm-diameter hole, covered by a grid of
80\% optical transparency, through which the electrons appearing
in the SFI were guided to an input window of a standard
channeltron VEU-6. The excitation region was surrounded by copper
shields, and all parts of the detection system were cooled to a
liquid-nitrogen temperature 77 K to decrease the influence of the
blackbody radiation. The laboratory magnetic field was compensated
for below 5 mGs in all directions by three pairs of Helmholtz
coils.

A microwave radiation tunable in the 53-78 GHz range was
introduced to the excitation region through a hole in a waveguide
connected to a standard backward-tube oscillator G4-142. It was
locked to a quartz synthesizer and operated in cw mode with a
linewidth below 10 kHz. The spectra of microwave transitions were
recorded by scanning the frequency of the synthesizer. Spatial
distribution of the microwave field was rather complex since it
was formed as arbitrary standing wave after multiply reflections
from surrounding copper surfaces that may be considered as a low-Q
(Q$\sim $100) microwave cavity. The point of laser excitation was
adjusted to a maximum of the standing wave following the procedure
described in our previous work \cite{29}.

The time evolution of the signals was as follows. At time $t = 0$
a laser pulse of 50 ns duration populated the 37S$_{1/2}$ state.
Microwave radiation tuned near the frequency of the one-photon
37S$_{1/2} \to $37P$_{1/2}$ or two-photon 37S$_{1/2} \to
$38S$_{1/2}$ transition was applied continuously. It populated the
final 37P$_{1/2}$\textit{} or 38S$_{1/2}$ state. The interaction
of atoms with radiation occurred until 2.8 $\mu $s, at which
moment a ramp of strong electric field with a rise-time of 2 $\mu
$s was applied for the SFI detection. The output pulses of the
channeltron appeared at 3.5-4.5 $\mu $s delay relative to the
laser pulse when the field reached critical values for the
different states. Final 37P$_{1/2}$ and 38S$_{1/2}$ states had
equal critical fields for the SFI, since in a strong electric
field their energies coincide, as can be seen from the Stark map
in Fig.3.

The SFI technique allows for detection of single Rydberg atoms
with probability close to 1 if electrons are detected by the
channeltron (our channeltron had typical gain about 10$^{8}$). A
total charge of the output pulses of the channeltron was
integrated separately for the lower and upper states of
transitions for each laser pulse, and measured by fast
Analog-to-Digital Converter (ADC). After amplification,
single-electron pulses had 400 mV average amplitudes that
fluctuated within 350-450 mV from pulse to pulse with a gaussian
distribution of amplitudes. Two-electron pulses had 800 mV average
amplitudes and fluctuated within 700-900 mV. Tree-electron pulses
were peaked at 1200 mV, etc. Selective determination of the total
number of Rydberg atoms was achieved up to five atoms, after which
value the multi-electron signals started to overlap. We thus were
able to distinguish up to 5 atoms and study the spectra of
multi-atom excitations at microwave transitions.

\subsection{Single-atom spectra}

The schemes of the one-photon 37S$_{1/2}\to $37P$_{1/2}$ and
two-photon 37S$_{1/2}\to $38S$_{1/2}$ microwave transitions in Na
atoms are shown in Fig.5. The relevant effective lifetimes are
about 60 $\mu $s for the S-states and 150 $\mu $s for the P-state,
so that spontaneous decay may be neglected during the interaction
time $\tau $=2.8 $\mu $s. In this approximation, single-atom
probabilities of transitions are calculated within the two- and
three-level models in a rotating-wave approximation for a square
microwave pulse. The evolution of populations of the upper state
37P$_{1/2}$ of the one-photon transition and 38S$_{1/2}$ of the
two-photon transition is described by the following formulas:

\begin{figure}
\includegraphics[scale=1]{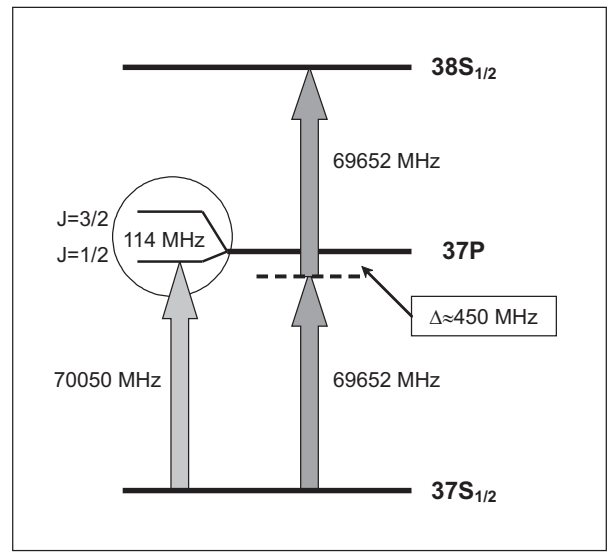}
\caption{\label{Fig5} Scheme of the one-photon 37S$_{1/2} \to
$37P$_{1/2}$ and two-photon 37S$_{1/2} \to $38S$_{1/2}$
transitions in sodium Rydberg atoms. }
\end{figure}

\begin{equation}
\label{eq11}
 \rho _{37P} = \frac{{\Omega _{1}^{2}} }{{\delta ^{2}
+ \Omega _{1}^{2} }}\sin^{2}\left( {\frac{{\tau} }{{2}}\sqrt
{\delta ^{2} + \Omega _{1}^{2}} } \right),
\end{equation}

\begin{equation}
\label{eq12}
 \rho _{38S} = \frac{{\Omega _{2}^{2}} }{{\left(
{\delta - \delta _{0}} \right)^{2} + \Omega _{2}^{2}}
}\sin^{2}\left( {\tau \sqrt {\left( {\delta - \delta _{0}}
\right)^{2} + \Omega _{2}^{2}} }  \right).
\end{equation}

\noindent
Here $\Omega _{1}=(d_{1} E/\hbar) $ is a Rabi frequency
of the 37S$_{1/2} \to $37P$_{1/2}$ transition (\textit{d}$_{1}$ is
a matrix element of the dipole moment, \textit{E} is the strength
of the microwave field), $\delta $ is a detuning from exact
resonance; $\Omega _{2} = d_{1} d_{2} E^{2}/\left( {4\hbar
^{2}\Delta}  \right)$ is a Rabi frequency of the 37S$_{1/2} \to
$38S$_{1/2}$ transition (\textit{d}$_{1}$, \textit{d}$_{2}$ are
matrix elements of the dipole moment for the 37S$ \to $37P and
37P$ \to $37S intermediate transitions, $\Delta $ is a detuning of
the virtual intermediate level of the two-photon transition from
the real 37P level), and $\delta _{0} = \left( {d_{1}^{2} -
d_{2}^{2}}  \right)E^{2}/\left( {8\hbar ^{2}\Delta}  \right)$ is a
power shift of the two-photon transition.

The experimental spectra obtained for the one- and two-photon
transitions are compared in Fig.6. The solid lines are theoretical
curves from Eqs.(\ref{eq11}) and (\ref{eq12}). In this experiment
we attempted to realize a NOT operation (inversion) in the center
of atomic line. The microwave intensity was adjusted to achieve
maximum population of the upper state in the center. It is seen
from Fig.6 that in the case of one-photon transition population of
the 37P$_{1/2}$ state does not exceed 70\% , and the shape of the
resonance substantially differs from the theoretical one. The
shape is closer to Lorentzian rather than to that predicted by
Eq.(\ref{eq11}). In the case of two-photon transition, population
of the upper 38S$_{1/2}$ state reaches 90\%, and the shape of the
resonance better agrees with theory. There are two reasons for
distortion of the experimental single-atom spectra.

\begin{figure}
\includegraphics[scale=1]{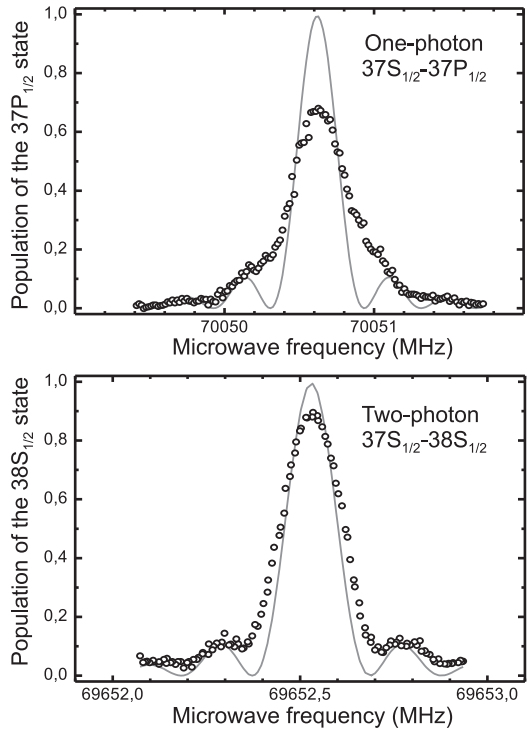}
\caption{\label{Fig6} Comparison of experimental records of the
spectrum of the one-photon 37S$_{1/2} \to $37P$_{1/2}$ transition
(upper picture) and two-photon 37S$_{1/2} \to $38S$_{1/2}$
transition (lower picture) with theoretical calculations (solid
lines).
 }
\end{figure}

First, atoms in the atomic beam have a Maxwell velocity
distribution with average velocity 600 m/s. This value corresponds
to 150 kHz Doppler broadening of the one-photon microwave
transitions at 70 GHz. The broadening is less important for
two-photon transitions, since photons in the standing wave may be
absorbed from two traveling waves propagating in opposite
direction, and the Doppler effect is partly compensated for. More
important is that effective time of interaction depends on the
atom velocity since atoms are passed through the nodes and
antinodes of the standing wave, so that maxima and minima in the
spectra may wash out.

Second, the spectra in Fig.6 were recorded in identical
experimental conditions except the difference in the microwave
intensity. We believe that better agreement with theory for the
two-photon transition, and disagreement for the one-photon
transition originate also from significant difference in the
microwave photon density required for the NOT operation. According
to Eqs.(\ref{eq11}) and (\ref{eq12}), inversion is achieved with
$\pi/2$ microwave pulses:

\begin{equation}
\label{eq13} \left\{ {{\begin{array}{*{20}l}
 {\Omega _{1} \tau /2 = \pi /2}, \hfill \\
 {\Omega _{2} \tau = \pi /2}. \hfill \\
\end{array}} } \right.
\end{equation}

These relations allow the calculation of \textit{E} value,
provided relevant dipole moments are known. Their radial parts
were calculated numerically following to Ref.\cite{26}, and found
to be 1460 a.u. for the 37S$ \to $37P transition and 1430 a.u. for
the 37P$ \to $38S transition. Angular parts were taken for the
transitions induced by linearly polarized microwave field, with
the fine-structure taken into account.

The average density of microwave photons at the antinode of a
standing wave was estimated from the following energy relation:
volume density of the energy of the electromagnetic field is

\begin{equation}
\label{eq14} \frac{{dW}}{{dV}} = \frac{{\varepsilon _{0}
E^{2}}}{{2}},
\end{equation}

\noindent where $\varepsilon _{0}$ is a dielectric constant.
Dividing this expression by $\hbar \omega $, we obtain a volume
density of photons \textit{dN}/\textit{dV} about 50 cm$^{-3}$ for
the one-photon 37S$_{1/2} \to $37P$_{1/2}$ transition, and about
10$^{5}$ cm$^{-3}$ for the two-photon 37S$_{1/2} \to $38S$_{1/2}$
transition.

\begin{figure*}
\includegraphics[scale=1]{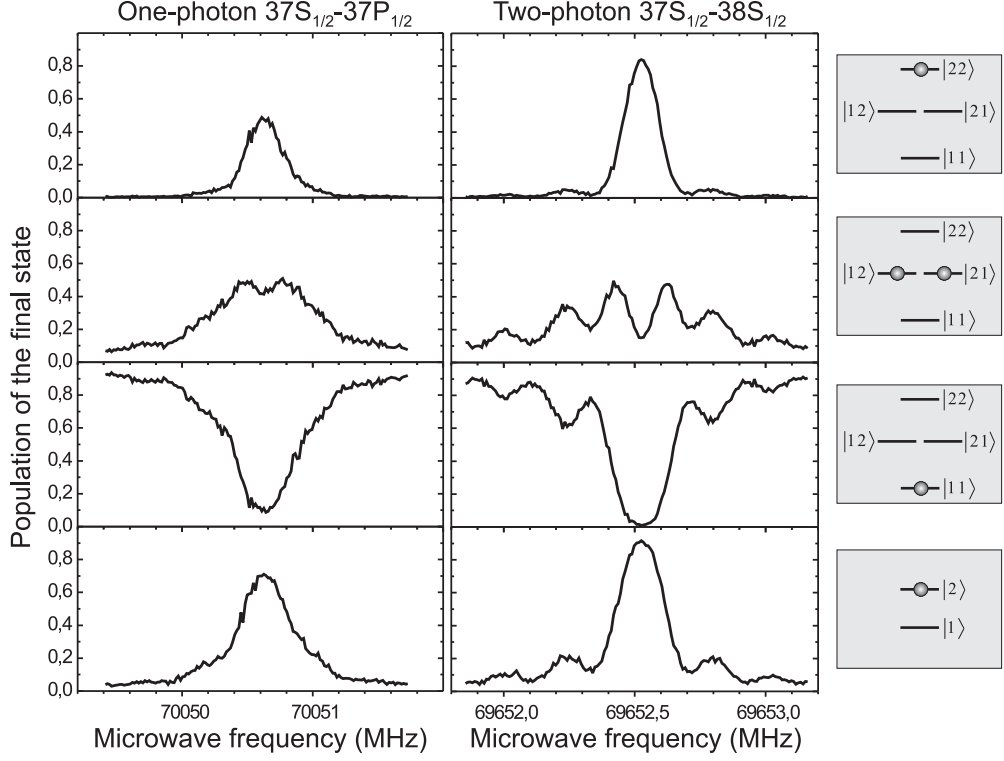}
\caption{\label{Fig7} Spectra of single-atom (lower trace) and
two-atom (upper traces) excitations of the one-photon 37S$_{1/2}
\to $37P$_{1/2}$ and two-photon 37S$_{1/2} \to $38S$_{1/2}$
transitions. The right-hand schemes indicate what final states of
a composite system are detected. State 1 is the initial
37S$_{1/2}$ state. State 2 is the final 37P$_{1/2}$ state (for the
one-photon transition) or 38S$_{1/2}$ state (for the two-photon
transition).
 }
\end{figure*}

Extremely low number of photons at the one-photon transition
requires huge attenuation of the output power of the microwave
oscillator. Even with such attenuation, we noticed that photons
leaking the oscillator through its case penetrated into the
interaction region through the holes in the detection system and
interfered with photons coming from the waveguide. This
interference caused instability in the spatial distribution of the
microwave field and also resulted in fluctuations of effective
interaction time. To some extent, this assumption was confirmed
experimentally when we decreased the time of interaction to 0.3
$\mu $s, and increased the microwave intensity in the waveguide by
three times. The one-photon resonance became wider, but in the
center of the line we were able to transfer about 80\% of
population to the upper state, and better agreement of the
spectrum with Eq.(\ref{eq11}) was found.

Obviously, the Doppler effect will be suppressed to a high degree
in the case of cold atoms trapped in the optical lattices, but
vibrational motion of atoms and fluctuations of laser intensity
and frequency may affect the precision of NOT operations in a way
similar to that in Fig.6.

\subsection{Multi-atom spectra}

Our detection system measured the total number of Rydberg atoms
and distribution of populations after each laser pulse. We
accumulated multi-atom signals for a short time (1-2 s) and
automatically sorted them by number of detected atoms and by
distribution of populations. No dipole-dipole interaction was
observed in this experiment since average distance between Rydberg
atoms was $\sim $100 $\mu $m. The experiments were made with of
one-photon 37S$_{1/2} \to $37P$_{1/2}$ and two-photon 37S$_{1/2}
\to $38S$_{1/2}$ transitions. We denote the initial 37S$_{1/2}$
state to be state 1, and the final states 37P$_{1/2}$ or
38S$_{1/2}$ to be state 2.

In the simplest case of two-atom excitations, an initial state is
$\left| {11} \right\rangle $, while four final states of two atoms
may be detected after interaction with the microwave field:
$\left| {11} \right\rangle $, $\left| {12} \right\rangle $,
$\left| {21} \right\rangle $, and $\left| {22} \right\rangle $.
Direct products of single-atom wavefunctions describe these states
if atoms do not interact with each other. The probabilities to
find two atoms after a measurement in a particular state may be
expressed as

\begin{equation}
\label{eq15} \left\{ {{\begin{array}{*{20}l}
 {\rho _{\left| {11} \right\rangle}  = \rho _{37S} \times \rho _{37S}}  \\
 {\rho _{\left| {12} \right\rangle}  = \rho _{\left| {21} \right\rangle}  =
\rho _{37S} \times \rho _{37P}}  \\
 {\rho _{\left| {22} \right\rangle}  = \rho _{37P} \times \rho _{37P}}  \\
\end{array}} } \right.
\end{equation}

\noindent for the one-photon transition, and

\begin{equation}
\label{eq16} \left\{ {{\begin{array}{*{20}l}
 {\rho _{\left| {11} \right\rangle}  = \rho _{37S} \times \rho _{37S}}  \\
 {\rho _{\left| {12} \right\rangle}  = \rho _{\left| {21} \right\rangle}  =
\rho _{37S} \times \rho _{38S}}  \\
 {\rho _{\left| {22} \right\rangle}  = \rho _{38S} \times \rho _{38S}}  \\
\end{array}} } \right.
\end{equation}

\noindent for the two-photon transition. Here $\rho _{37P}$ and
$\rho _{38S}$ are the probabilities to find single atom in a final
state given by Eqs.(\ref{eq11}) and (\ref{eq12}), and $\rho
_{37S}$=(1-$\rho _{37P}$) or $\rho _{37S}$=(1-$\rho _{38S}$) are
probabilities of the single atom to remain in the 37S state.

In Fig.7 we present the experimental records of the two-atom
spectra (three upper traces), and compare them with a single-atom
spectrum (lower trace). Microwave intensity was adjusted to
provide maximum population transfer to the upper state. The
right-hand schemes show what final states of the composite system
of two atoms were detected. The comparison of the spectra with
calculations by Eq.(\ref{eq16}) revealed good agreement for the
two-photon transition, confirming that we indeed detected only two
atoms. Agreement was much worse for the one-photon transition due
to the reasons discussed in the previous subsection.

All two-atom spectra are different from that of a single atom. We
note two important features: (i) probability of excitation of only
one of two atoms has a narrow dip in the line center due to large
probability to excite both atoms, and (ii) probability of
excitation of both atoms has narrower spectrum than single-atom
spectrum. This may have applications for high-resolution
spectroscopy as a method of narrowing of atomic lines. For
example, in the case of \textit{N} Rydberg atoms the probability
to transfer all atoms to the upper state 2 is

\begin{equation}
\label{eq17} \rho _{\left| {222......2} \right\rangle}  = \left(
{\rho _{38S}} \right)^{N}
\end{equation}

\noindent for the two-photon transition. If this final state is
selectively detected in the atomic ensemble, as in our experiment,
observed spectrum of full excitation will be much narrower than
single atom spectrum. The width of multi-atom resonances may be
estimated by substituting the Lorentzian pre-factor of
Eq.(\ref{eq12}) to Eq.(\ref{eq17}). The full width at half maximum
is

\begin{equation}
\label{eq18} \Delta \,\omega _{N} = 2\Omega _{2} \sqrt
{\sqrt[{N}]{{2}} - 1}\quad .
\end{equation}

\noindent
 In the limit of \textit{N$\gg$}1, one finds the
relationship between the widths of the single-atom and multi-atom
resonances:

\begin{equation}
\label{eq19} \frac{{\Delta \,\omega _{1}} }{{\Delta \,\omega _{N}}
} \approx \sqrt {\frac{{N}}{{\ln 2}}} \quad .
\end{equation}

\noindent
For example, the resonance becomes 2.7 times narrower
with \textit{N}=5.

\begin{figure}
\includegraphics[scale=1]{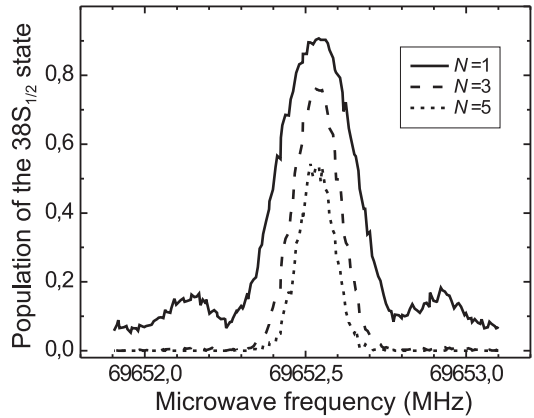}
\caption{\label{Fig8} Spectra of full multi-atom excitations of
the two-photon transition 37S$_{1/2}\to $38S$_{1/2}$ for various
number \textit{N} of Rydberg atoms.
 }
\end{figure}

In Fig.8 the spectra of full multi-atom excitations observed at
the two-photon transition are presented. The spectra of full
excitations for 1, 3, and 5 atoms display that multi-atom
resonances are indeed much narrower than single-atom ones, but
their amplitudes become smaller with growing \textit{ N}. We
conclude that ultimate resolution of this method of narrowing of
atomic lines is limited by the ability to distinguish full
excitation of the atomic ensemble from the partial one, and by the
ability to transfer populations of all atoms to the upper states,
otherwise amplitude of the multi-atom resonance would rapidly
decrease with growing $N$.

Although our experiments were made with atomic beam and no
dipole-dipole interaction was observed, obtained spectra revealed
the main features of single- and multi-atom excitations. We expect
that similar spectra would be observed, e.g., at the laser
excitation of an ensemble of Rydberg atoms from the ground state.

Now we can also understand how the dipole-dipole interaction may
affect the spectra. In the case of two Rydberg atoms and
one-photon \textit{n}S$_{1/2} \to $\textit{n}P$_{1/2}$ microwave
transition, it will change the frequencies of transitions from the
initial $\left| {\mathrm{SS}} \right\rangle $ state to the
strongly interacting $\left| {\mathrm{SP}} \right\rangle $ and
$\left| {\mathrm{PS}} \right\rangle $ states, while the frequency
of the transition to $\left| {\mathrm{PP}} \right\rangle $ state
will be unaffected but its probability will decrease. An
experiment on the observation of the dipole-dipole interaction
between few close Rydberg atoms excited in extremely small volume
of a dense atomic beam is in progress now.

\section{SUMMARY}

We have analyzed the applicability of Rydberg atoms to quantum
computers and defined the parameters of Rydberg atoms and
experimental arrangements that are most suitable to realize a
quantum phase gate in optical lattices of cold alkali atoms. The
principal quantum number \textit{n}$\sim $30, low orbital moments
(\textit{L}$\sim $0, 1, 2), $\sim $5 $\mu $m interatomic distance,
50 ns exciting laser pulses, and $\sim $500 ns time of
dipole-dipole interaction seem to be optimal. Modeling experiments
on microwave spectroscopy of a few Na Rydberg atoms in the atomic
beam revealed some important features in the spectra of single-
and multi-atom excitation for the one- and two-photon transitions.
The quantum NOT operation with single atoms was found to be
affected by the Doppler effect and fluctuations of the microwave
field. The spectrum of complete excitation of several Rydberg
atoms was much narrower than that of single atom. This effect may
be applied to high-resolution spectroscopy for the narrowing of
atomic lines. Similar spectra of single- and multi-atom
excitations will be observed at the laser excitation of an
ensemble of ground-state atoms to Rydberg states.

\begin{acknowledgments}
This work was supported by the Russian Foundation for Basic
Research, Grant Nos. 02-02-16323, 03-02-06867, and by INTAS, Grant
No.2001-155.
\end{acknowledgments}


\begin{thebibliography}{10}

\bibitem{1}
D.~P.~DiVincenzo, Science \textbf{270}, 255 (1995); J.~Gruska,
\textit{Quantum Computing} (McGraw-Hill, London, 1999);
A.~M.~Steane, Rep. Prog. Phys. 61, 117(1998).

\bibitem{2} C.~H.~Bennett and D.~P.~DiVincenzo, Nature (London)
\textbf{404}, 247 (2000).

\bibitem{3} H.~J.~Metcalf and P.~van~der~Straten, \textit{Laser
Cooling and Trapping} (Springer, New York, 1999); R.~Grimm,
M.~Weidem$\ddot{\mathrm{u}}$ller, and Yu.~B.~Ovchinnikov, Adv. At.
Mol. Opt. Phys. \textbf{42}, 95 (2000).

\bibitem{4} R.~Dumke, M.~Volk, T.~Muther, F.~B.~J.~Buchkremer,
G.~Birkl, and W.~Ertmer, Phys. Rev. Lett. \textbf{89}, 097903
(2002).

\bibitem{5} N.~Schlosser, G.~Reymond, I.~Protsenko, and
P.~Granier, Nature (London) \textbf{411}, 1024 (2001).

\bibitem{6} N.~Schlosser, G.~Reymond, and P.~Grangier, Phys. Rev.
Lett. \textbf{89}, 023005 (2002).

\bibitem{7} R.~P.~Feynman, Opt. News \textbf{11}, 11 (1985).

\bibitem{8} G.~K.~Brennen, C.~M.~Caves, P.~S.~Jessen, and
I.~H.~Deutsch, Phys. Rev. Lett. \textbf{82}, 1060 (1999);
G.~K.~Brennen, I.~H.~Deutsch, and P.~S.~Jessen, Phys. Rev. A
\textbf{61}, 062309 (2000).

\bibitem{9}D.~Jaksch, J.~I.~Cirac, P.~Zoller, S.~L.~Rolston,
R.~C$\hat{\mathrm{o}}$t$\acute{\mathrm{e}}$, and M.~D.~Lukin,
Phys. Rev. Lett. \textbf{85}, 2208 (2000).

\bibitem{10}M.~D.~Lukin, M.~Fleischhauer, R.~C$\hat{\mathrm{o}}$t$\acute{\mathrm{e}}$, L.~M.~Duan,
D.~Jaksch, J.~I.~Cirac, and P.~Zoller, Phys. Rev. Lett.
\textbf{87}, 037901 (2001).

\bibitem{11}\textit{ Rydberg States of Atoms and Molecules}, ed.
by R.~F.~Stebbings and F.~B.~Dunning (Cambridge University Press,
Cambridge, 1983); T.~F.~Gallagher, \textit{Rydberg Atoms}
(Cambridge University Press, Cambridge, 1994).

\bibitem{12}I.~E.~Protsenko, G.~Reymond, N.~Schlosser, and
P.~Grangier, Phys. Rev. A \textbf{65}, 052301 (2002).

\bibitem{13}M.~S.~Safronova , C.~J.~Williams,  and C.~W.~Clark, Phys. Rev. A
 \textbf{67}, 040303(R) (2003).

\bibitem{14}R.~G.~Unanyan and M.~Fleischhauer, Phys. Rev. A
\textbf{66}, 032109 (2002).

\bibitem{15}C.~Boisseau , I.~Simboten  and R.~C$\hat{\mathrm{o}}$t$\acute{\mathrm{e}}$, Phys. Rev. Lett.
\textbf{88}, 133004 (2002); S.~M.~Farooqi, D.~Tong, S.~Krishnan,
J.~Stanojevic, Y.~P.~Zhang, J.~R.~Ensher, A.~S.~Estrin,
C.~Boisseau, R.~C$\hat{\mathrm{o}}$t$\acute{\mathrm{e}}$,
E.~E.~Eyler,  and P.~ L.~Gould, Phys. Rev. Lett. \textbf{91},
183002 (2003).

\bibitem{16} J.~M.~Raimond, G.~Vitrant,  and S.~Haroche,
J. Phys. B \textbf{14}, L655 (1981).

\bibitem{17}D.~Tong, S.~M.~Farooqi, J.~Stanojevic, S.~Krishnan, Y.~P.~Zhang, R.~C$\hat{\mathrm{o}}$t$\acute{\mathrm{e}}$,
E.~E.~Eyler,  and P.~L.~Gould, arXiv:physics/0402113 (2004);
K.~Singer, M.~Reetz-Lamour, T.~Amthor, L.~G.~Marcassa,  and
M.~Weidem$\ddot{\mathrm{u}}$ller, arXiv:physics/0404075 (2004).

\bibitem{18} W.~R.~Anderson, J.~R.~Veale, and T.~F.~Gallagher,
Phys. Rev. Lett. \textbf{80}, 249 (1998); I.~Mourachko,
D.~Comparat, F.~de~Tomasi, A.~Fioretti, P.~Nosbaum, V.~M.~Akulin,
and P.~Pillet, Phys. Rev. Lett. \textbf{80}, 253 (1998).

\bibitem{19} K.~Afrousheh, Z.~Bohlouli, D.~Vagale, A.~Mugford,
M.~Fedorov, and J.~D.~D.~Martin,  Submitted to Phys. Rev. Lett.
(2004).

\bibitem{20} H.~A.~Bethe and E.~E.~Salpeter, \textit{Quantum
Mechanics of One- and Two-Electron Atoms} (Springer, Berlin,
1957).

\bibitem{21} J.~Neukammer, H.~Rinneberg, K.~Vietzke, A.~Konig,
H.~Hieronymus, M.~Kohl, H.~-J.~Grabka, and G.~Wunner, Phys. Rev.
Lett. \textbf{59}, 2947 (1987).

\bibitem{22} S.~Osnaghi, P.~Bertet, A.~Auffeves, P.~Maioli,
M.~Brune, J.~M.~Raimond, and S.~Haroche, Phys. Rev. Lett
\textbf{87}, 037902 (2001).

\bibitem{23} I.~M.~Beterov and I.~I.~Ryabtsev, JETP Lett.
\textbf{69}, 448 (1999).

\bibitem{24} S.~Lloyd, Phys. Rev. Lett. \textbf{75}, 346 (1995).

\bibitem{25} A.~Rauschenbeutel, G.~Nogues, S.~Osnaghi, P.~Bertet,
M.~Brune, J.~M.~Raimond, and S.~Haroche, Phys. Rev. Lett.
\textbf{83}, 5166 (1999).

\bibitem{26} M.~L.~Zimmerman, M.~G.~Littman, M.~M.~Kash, and
D.~Kleppner, Phys. Rev. A \textbf{20}, 2251 (1979).

\bibitem{27} A.~R.~Edmonds, J.~Picart, N.~Tran-Minh, and
R.~Pullen, J. Phys. B \textbf{12}, 2781, (1979).

\bibitem{28} J.~M.~Raimond, M.~Brune, and S.~Haroche, Rev. Mod.
Phys. \textbf{73}, 565 (2001).

\bibitem{29} I.~I.~Ryabtsev and D.~B.~Tretyakov, Phys. Rev. A
\textbf{64}, 033413 (2001).

\bibitem{30} I.~I.~Ryabtsev and I.~M.~Beterov, Phys.
Rev. A \textbf{61}, 063414 (2000).

\bibitem{31} I.~I.~Ryabtsev, D.~B.~Tretyakov, and I.~I.~Beterov,
J. Phys. B \textbf{36}, 297 (2003).

\bibitem{32} C.~E.~Theodosiou, Phys. Rev. A \textbf{30},
2881 (1984).

\end{thebibliography}
\end{document}